\global\newif\ifpreprint
\global\preprintfalse

\ifpreprint
     \documentstyle[aps,epsfig,preprint,endfloat]{revtex}
\else
     \documentstyle[aps,epsfig,multicol]{revtex}
\fi

\begin{document}

\title{Band structure of MoS$_2$, MoSe$_2$, and $\alpha$-MoTe$_2$:
  Angle-resolved photoelectron spectroscopy and
  {\em ab-initio}\/ calculations}

\author{Th.~B{\"o}ker, R.~Severin, A.~M{\"u}ller, C.~Janowitz,
  and R.~Manzke}
\address{%
  Institut f{\"u}r Physik,
  Humboldt-Universit{\"a}t zu Berlin,
  Invalidenstra{\ss}e 110, 10115 Berlin}

\author{D.~Vo{\ss}, P.~Kr{\"u}ger, A.~Mazur, and J.~Pollmann} 
\address{%
  Institut f{\"u}r Festk{\"o}rpertheorie,
  Universit{\"a}t M{\"u}nster, 
  Wilhelm-Klemm-Stra{\ss}e 10, 48149 M{\"u}nster}

\maketitle
\vspace*{1.0cm}

\bibliographystyle{prsty}


\begin{abstract}
In this work the complete valence-band structure of the molybdenum 
dichalcogenides MoS$_2$, MoSe$_2$, and $\alpha$-MoTe$_2$ is presented and 
discussed in comparison. The valence bands have been studied 
using both angle-resolved photoelectron spectroscopy (ARPES) with
synchrotron radiation, as well as, {\em ab-initio}\/ band-structure calculations.
The ARPES measurements have been carried out in the constant-final-state (CFS)
mode. The results of the calculations show in general very good agreement with 
the experimentally determined valence-band structures allowing for a clear 
identification of the observed features.
The dispersion of the valence bands as a function of the perpendicular 
component $\vec{k}_\perp$ of the wave vector reveals a decreasing
three-dimensional character from MoS$_2$ to $\alpha$-MoTe$_2$ which
is attributed to an increasing interlayer distance in the three compounds.
The effect of this $\vec{k}_\perp$ dispersion on the determination of the 
exact dispersion of the individual states as a function of $\vec{k}_\parallel$ 
is discussed. By performing ARPES in the CFS mode the $\vec{k}_\parallel$-component 
for off-normal emission spectra can be determined.
The corresponding $\vec{k}_\perp$-value is obtained from the symmetry
of the spectra along the $\Gamma$A, KH, and ML line, respectively.
\end{abstract}

\draft

\pacs{\\{\em Keywords}\/: 71.20; 79.60; 31.15.A}


\ifpreprint
\else
  \begin{multicols}{2}
\fi

\section{Introduction}

The molybdenum dichalcogenides belong to the large family of layered
transition metal dichalcogenides whose crystal structure results from
the stacking of sheets of hexagonally packed atoms.
It consists of weakly coupled sandwich layers \mbox{X-Mo-X} 
in which a Mo-atom layer is enclosed within two chalcogen layers 
(X = S, Se, Te). All three MoX$_2$ compounds crystallize in the 2H$_{\rm b}$
structure and MoS$_2$, as well as, MoSe$_2$  also crystallize in the 3R structure. 
In both of these polytype structures the coordination of 
the metal atoms is trigonal prismatic. In this work, 2H$_{\rm b}$ samples 
are investigated. For MoTe$_2$ a phase transition from the semiconducting
2H$_{\rm b}$-type $\alpha$-MoTe$_2$ to a metallic polytype, $\beta$-MoTe$_2$, 
with a monoclinic structure having a distorted octahedral coordination
is known\cite{LieTer77,HlMaPr96,Gmelin95}.
The 2H$_{\rm b}$-polytype, in the following assigned as MoTe$_2$, is
stable below $\approx815^{\circ}$C.

The MoX$_2$ compounds are indirect semiconductors with indirect (direct) band
gaps of 1.29 (1.78)~eV for MoS$_2$, 1.10 (1.42)~eV for MoSe$_2$, 
and 1.00 (1.10)~eV for MoTe$_2$ \cite{Gmelin95}.
Since the optical band gaps are matching well with the solar
spectrum these materials are used for electrodes in high efficiency
photo-electrochemical (PEC) cells \cite{AbHoBa82}. In the
molybdenum dichalcogenides the phototransitions
involve non-bonding $d$ orbitals of Mo atoms. Therefore, these materials 
can be expected to resist hole-induced corrosion \cite{KeCoBo90}.
Despite of the major importance of the MoX$_2$ compounds in solar cell 
production, to date no comparative band structure determination from
experiment and theory has been reported. 

In general, it is expected that the gross features of the band structures 
of all three dichalcogenides are quite similar but, in addition, also some 
characteristic differences are to be expected.
Several band structure calculations were performed so far which were
used to explain experimental results from absorption measurements 
\cite{BeKnLi72,BrMuYo72,Evans76,Liang73} and early photoemission work 
\cite{CoHaDi87,CoHaGr87,MaBoCa87,FiMcMc92}. From
the literature about these materials no consistent picture emerges
(see e.g.\  \cite{LanBor94} for a review, and references therein). Therefore, 
we address in this work the electronic band structure of all three binary 
dichalcogenides on equal footing by ARPES measurements and 
by 'state-of-the-art' {\em ab-initio}\/ band structure calculations.
In experiment, high-resolution photoelectron spectrometers (with respect to 
emission angle and energy) are used.  In addition,
the constant-final-state (CFS) mode is applied to obtain data which shows low
$\vec{k}_\perp$ dispersion. As to the theory, the electronic properties 
of all three molybdenum dichalcogenides are calculated within density functional 
theory (DFT) employing the local density approximation (LDA)~\cite{Hoh64} 
and spin-orbit interaction is included.

The present photoemission data and {\it ab-initio} results constitute  
the first comprehensive picture of the electronic valence-band structure 
of all three molybdenum dichalcogenides.

\section{Crystal structure}

The 2H$_{\rm b}$-polytype of the crystal structure of the MoX$_2$ layer compounds
is characterized by a stacking sequence /AbA BaB/ (A, B: chalcogen atom layers;
a, b: Mo atom layers) \cite{WilYof69}. A unit cell and the corresponding first 
Brillouin zone are displayed in Fig.\ref{fig:crystal}. In this structure, a 
Mo atom layer is sandwiched between two chalcogen layers. The two X-Mo-X 
sandwich layers per unit cell are laterally displaced with respect to each 
other so that the Mo atoms of the upper sandwich layer are directly above the 
chalcogen atoms of the lower sandwich layer in the unit cell and vice versa.
The sandwich layers are coupled only by weak van der Waals 
forces  giving rise to the quasi-two-dimensional character of the electronic 
structure. 

The lattice parameters of the direct lattice $a$, $c$, $w$ and $z$ are indicated
in Fig.\ref{fig:crystal}. In this work, we use averages of the $a$ and $c$  
values, as  reported in the literature from different experiments 
\cite{Gmelin95,LanBor94,WilYof69}. The thickness of a sandwich layer
is taken to be 2\,$z$ with $z=0.129\cdot c$. Therefore, an interlayer
distance $w = c/2-2\,z$ = $0.242\cdot c$ follows. In Tab.~\ref{tab:strcdata} 
the used average experimental structure parameters, the resulting $c/a$ ratios,
and the calculated reciprocal lengths of the most important high-symmetry 
lines (see Fig.~\ref{fig:crystal}~(b)) are given.  
In addition, the covalent radii of the chalcogen atoms and the ionic radii of 
the twofold negatively charged chalcogen ions are given. In the dichalcogenide 
crystals, the actual charge of the chalcogens is between the two limiting 
values of 0$e$ and -2$e$ as a simple population analysis of the calulated 
density of states shows. It is obvious from Tab.~\ref{tab:strcdata} that
the interlayer distance $w$, as well as, the lattice parameters $a$ and
$c$ increase from MoS$_2$ to MoTe$_2$ mainly due to the increasing ionic
radius of the chalcogen ions, while the $c/a$ ratio remains nearly constant.

\begin{figure} 
\begin{minipage}{\linewidth} 
       \centerline{\epsfig{file=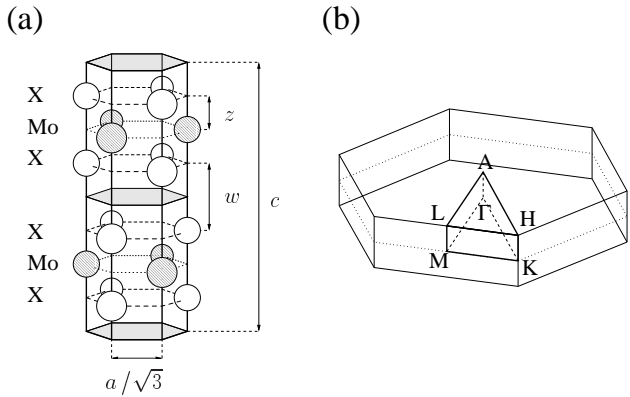, angle=0, width=\linewidth}} 
       \caption{Crystal structure (a) and the first Brillouin zone  
                (b) of the 2H$_{\rm b}$ structure of the MoX$_2$.  
        \label{fig:crystal}} 
\end{minipage} 
\end{figure} 

\begin{minipage}{\linewidth}\noindent 
\begin{table} 
  \caption[]{Structural parameters of the 2H$_{\rm b}$ polytype of the 
    molybdenum dichalcogenides for the direct and reciprocal lattice. 
    The parameter $z$, i.e.\  half the thickness of a sandwich layer, is 
    given by $z=0.129\cdot c$. The size of the v.\,d.\,Waals gap $w$ is 
    calculated as $c/2-2\,z$. 
    Furthermore the size of the Brillouin zone is determined by the respective  
    $k_\parallel$ values. 
    In addition, the covalent and ionic radii of the chalcogen 
    atoms are given. 
    The corresponding radii of molybdenum are Mo: 1.40\,\AA\/ and 
    Mo$^{4+}$: 0.70\,\AA, respectively. 
    \label{tab:strcdata}} 
\begin{center} 
\begin{tabular}{l|r|r|r} 
                      & {\bf MoS$_2$}  & {\bf MoSe$_2$} %
  & {\bf MoTe$_2$}  \\ \hline 
$a~[{\mbox{\AA}}]$    & 3.160          & 3.299          %
  & 3.522          \\ 
$c~[{\mbox{\AA}}]$    & 12.294         & 12.938         %
  & 13.968         \\ %
  \hline 
$2\,z~[{\mbox{\AA}}]$ & 3.172          & 3.338          %
  & 3.604          \\ 
$w~[{\mbox{\AA}}]$    & 2.975          & 3.131          %
  & 3.380          \\ 
$c/a$                 & 3.891          & 3.922          %
  & 3.966 \\ %
  \hline 
$\overline{\Gamma \mbox{A}}~[{\mbox{\AA}^{-1}}]$ & 0.255 & 0.243 %
  & 0.225 \\ 
$\overline{\Gamma \mbox{M}}~[{\mbox{\AA}^{-1}}]$ & 1.148 & 1.099 %
  & 1.030 \\ 
$\overline{\Gamma \mbox{K}}~[{\mbox{\AA}^{-1}}]$ & 1.325 & 1.270 %
  & 1.189 \\ 
$\overline{\rm MK}~[{\mbox{\AA}^{-1}}]$          & 0.662 & 0.636 %
  & 0.594 \\ 
  \hline 
$r_{cov}~[\mbox{\AA}]$ & S: 1.04       & Se: 1.17   %
  & Te: 1.37  \\ 
$r_{ion}~[\mbox{\AA}]$  & S$^{2-}$: 1.84& Se$^{2-}$: 1.98 %
  & Te$^{2-}$: 2.21 
\end{tabular} 
\end{center} 
\end{table} 
\end{minipage}

\section{Experimental approach}

\subsection{Sample preparation and experimental setup}

Naturally grown molybdenite (MoS$_2$) was used for this work. Excellent
samples with sizes up to $3\times 3$~mm$^2$ have been extracted from 
a larger piece. Single crystals of MoSe$_2$ and $\alpha$-MoTe$_2$
were grown using the chemical vapor transport method (CVT) where 
bromine was used as the transport agent. 

Coarse orientation of the samples was achieved by LEED-patterns which
were recorded under ultrahigh-vacuum conditions before cleavage. From 
the sharpness of the diffraction patterns it can be concluded 
that the single crystals are of high quality. Next, a 
small aluminum post was glued on top of the samples and they were
inserted into the photoemission chamber. After preadjustment, the
crystals were cleaved {\it in situ}\/ by knocking off the aluminum
posts. This was done at room temperature under ultrahigh-vacuum
conditions (p$\leq2\times10^{-10}$~mbar) 
to minimize surface degradation or contamination. Immediately 
after the cleavage a reference spectrum was recorded to check the 
surface quality from time to time. Even after up to fourty hours of 
data acquisition at room temperature no significant changes were
observed. This is due to the crystals cleaving at the van der Waals gap 
resulting in a rather inert surface.

The measurements have been performed at the synchrotron radiation
centers HASYLAB in Hamburg, Germany, and at BESSY-I in Berlin, Germany. 
The typical photon energy range of $h\nu=9\dots30$~eV and a 
monochromator resolution of below 90~meV was used. The
photoelectron spectrometers are hemispherical deflection analyzers
mounted on two-axis goniometers. These are the WESPHOA-III 
station at the HONORMI beamline (W3.2) at HASYLAB and the new
photoemission station AR 65 at the 3m-NIM1 and 2m-SEYA beamlines of
BESSY-I \cite{JaMuPl99}. For the measurements presented here, 
a monochromator resolution of 30-90~meV and an analyzer resolution 
of 50~meV were chosen, yielding an overall resolution of less than
100~meV. This is sufficient since all measurements were performed at 
room temperature. The maximum acceptance angle of the photoelectrons was
$\approx \Delta \vartheta \pm 0.8^\circ$.

\subsection{The CFS mode in photoemission}
Photoemission spectroscopy can be performed in several different modes
to determine the occupied bandstructure. Most common is the
energy distribution curve (EDC) mode in which the photon energy 
is held constant and the detected kinetic energy is varied. With the 
emission angle $\vartheta$ (with respect to the surface normal) and 
the detected kinetic energy E$_{\rm kin}$ of the photoelectrons the 
wave-vector component parallel to the planes is determined by the 
dispersion relation:

\begin{equation}
|\vec{k}_\parallel|=
\frac{1}{\hbar}\sqrt{2m_eE_{\rm kin}}\cdot\sin\vartheta .
\label{eq:k_par}
\end{equation}

The $\vec{k}_\perp$-component is not conserved since the electrons travel 
through the crystal surface. If the final state band can be approximated 
by a free electron parabola starting at the inner potential V$_0$, the 
following equation can be deduced:

\begin{equation}
|\vec{k}_\perp|=
\frac{1}{\hbar}\sqrt{2m_e(E_{\rm kin}\cdot\cos^2{\vartheta}+V_0)}
\label{eq:k_perp}
\end{equation}

Obviously, $\vec{k}_\perp$ has 
to be larger than $\frac{1}{\hbar}\sqrt{2m_eV_0}$ in order to be able to observe 
photoemitted electrons outside the sample. A key question for the analysis  
of the experimental data is whether this approximation can be used for the 
molybdenum dichalcogenides 
in the final state energy range used in this work. Recently Strocov 
{\it et al.}\/~\cite{StStNi98,StStNi96} have claimed that this approximation 
might not be suitable for two-dimensional layered transition-metal dichalcogenides.
We will come back to this point in section~\ref{sec:final-states}. 
In addition, Strocov {\it et al.}\/ found for TiS$_2$ and VSe$_2$ that 
the inner potential $V_0$ is dependent on $\vec{k}$ for these materials.
In the present study of the molybdenum dichalcogenides, $V_0$ has been 
determined by LEED-I/V curves where intial electron energies between 50...250~eV have
been used. For $V_0$ we have determined $12.0\pm2.3$ eV in MoS$_2$, $14.5\pm1.5$ eV 
in MoSe$_2$ and $16.1\pm1$ eV in MoTe$_2$. An error of about 1.5~eV can be assumed 
for the determination 
of $V_0$ by the LEED-I/V curve technique. Larger errors result from the statistical 
spread over several samples which have been investigated.
For normal emission the critical points can be found accurately from 
the symmetry of the dispersion of the uppermost band along 
$\overline{\Gamma {\rm A} \Gamma}$. By this procedure the experimental band
structure at the high symmetry points can directly be compared to the corresponding 
calculated band structure.

The constant-final-state mode has some advantages with respect to the
EDC mode. For example, Eqs.~\ref{eq:k_par}~and~\ref{eq:k_perp} show that 
both $\vec{k}$-components depend on the kinetic energy. Therefore, the
emission features in an EDC spectrum are affected by a dispersion in 
both directions. For a strongly dispersing band the corresponding peaks
are asymmetric within the EDC spectra. This can be avoided in the CFS 
mode, where both $\vec{k}$-components ($\vec{k}_\parallel$ and $\vec{k}_\perp$)
are kept constant. A spectrum at a new value of e.g.\  $\vec{k}_\parallel$ 
is adjusted by choosing a couple (E$_{\rm kin},\vartheta$) for which 
$\vec{k}_\perp$ remains constant. The range of these values is limited by 
the monochromator energy range so that not all high-symmetry directions 
are accessible. Another advantage of the CFS mode lies in the fact that 
the detected kinetic energy at the 
analyzer is kept constant, too. Since the transmission function increases for 
lower $E_{kin}$ values an EDC spectrum typically reveals an increased
background towards the low energy region, which cannot be determined
exactly. In the CFS mode, only the characteristics of the monochromator
(i.~e.\  photon flux per photon energy) influences the background of
each spectrum. But this can be determined exactly, e.g., by measuring 
the current of a reference gold-mesh which is positioned in the photon beam.

\subsection{Data analysis technique}
Each spectrum measured in the present work has been modeled by a curve 
fitting program based on the Levenberg-Marquardt algorithm \cite{PrTeVe92}. 
The model function is created by convoluting a sum of peaks plus 
Shirley-background \cite{Shirle72} with the Gaussian shaped spectrometer
function. Many emission features with larger binding energy reveal a 
Lorentzian line shape which is in agreement with the physical nature of
the electronic states. A typical full width at half maximum of
$200\dots400$~meV
indicates that the intrinsic line shape cannot be influenced by the 
spectrometer function since the latter was always chosen to be lower 
than 100~meV in width (i.e., analyzer plus monochromator resolution).
Since in the CFS mode the whole spectrum belongs to one specific 
$\vec{k}$ vector, the line shape cannot be influenced by dispersion effects 
which may occur in EDC spectra. 

\section{Band-structure calculations}   

The LDA-DFT calculations have been carried out employing the norm-conserving, 
nonlocal pseudopotentials of Bachelet, Hamann, and Schl{\"u}ter\cite{BHS82}.
The exchange-correlation energy was taken into account using the Ceperley%
-Alder\cite{Cep80} form as parametrized by Perdew and Zunger\cite{Per81}.
As a basis to represent the wavefunctions, we use 120
Gaussian orbitals of $s$, $p$, $d$, and $s^\ast$ symmetry {\it per
sandwich layer} and spin. At each atomic position, 40 orbitals are localized.
The decay constants of the Gaussians employed are
\{0.17, 0.45, 1.18, 2.00\} for Mo, \{0.17, 0.45, 0.90, 1.70\} for S,
\{0.17, 0.45, 0.90, 1.80\} for Se, and \{0.17, 0.43, 1.05, 2.60\} for Te
(in atomic units). 
A real space mesh with linear spacings of about 0.2\,\AA\ is used for the
representation of the charge density and the
potential. The spin-orbit interaction is treated in an on-site
approximation, i.e.,  only integrals with the same location of the
Gaussian orbitals and the spin-orbit potential are taken into account.
Brillouin-zone integrations have been carried out using 12 special
{\bf k}-points in the irreducible part of the Brillouin zone.

In our calculations we find, that a number of electronic features depend 
sensitively on the lattice parameters. Therefore, we have used in the 
calculations reported in this work a value of $z$ = $0.129\cdot c$ for half 
the diameter of the sandwich layers as obtained in experiment in order  
to arrive at a more meaningful comparison between theoretical and 
experimental results. Our calculations of $z$ by total energy minimization
(with $a$ and $c$ being held constant) yield  $0.125\cdot c$ for MoS$_2$, 
$0.128\cdot c$ for MoSe$_2$, and $0.127\cdot c$ for MoTe$_2$, i.e.,
values which are somewhat lower than the experimental values.

\section{Results and discussion}

\subsection{Final states}
\label{sec:final-states}

As mentioned above, the knowledge of the final state and the
corresponding $\vec{k}_\perp$ component is very important when
the CFS mode is applied in photoemission. Eq.~\ref{eq:k_perp} assumes 
free-electron-like final-state bands which in general, as claimed by 
Strocov {\it et al.} \cite{StStNi98}, might not be a good approximation for 
layered transition-metal dichalcogenides.  As a result from their study 
of TiS$_2$ and VSe$_2$, the authors argued that the complicated dispersion 
of the final state bands affects the band mapping in a way that the 
$\vec{k}_\perp$ component remains undetermined. In contrast, it can be  
expected for quasi-two-dimensional materials and especially for the MoX$_2$ 
compounds that the $E(\vec{k}_\perp)$ dependence is significantly smaller than 
the $E(\vec{k}_\parallel)$ dependence.

To determine experimentally the critical points at the zone edge a number of
CFS spectra have been recorded. In particular, CFS spectra with 
$\vec{k}_\parallel$ vectors along the $\Gamma$K direction at four different 
$\vec{k}_\perp$ values corresponding to (0.0 $\pi$/c, -0.25 $\pi$/c, 0.50 $\pi$/c 
and 1.0 $\pi$/c) were recorded and analyzed. The first of these
four values corresponds to spectra taken along the $\Gamma$K and the last  
along the AH direction in the Brillouin zone.
The experiments were carried out for MoS$_2$ at the HONORMI beamline 
in HASYLAB (DESY, Hamburg) where a photon energy range of $h\nu=10\dots30$~eV 
can be used. MoS$_2$ was chosen since this material is expected to reveal 
the most pronounced three-dimensional character in the band structure.
Nevertheless, the respective dispersion in most cases was found to be very 
small, as compared to the dispersion of the bands along the $\vec{k}_\parallel$ 
in-plane directions. Within the available photon energy range it was 
possible to record CFS series along the $\Gamma$K\,(AH) directions at
four different $\vec{k}_\perp$ values. 

\begin{figure}\noindent 
\begin{minipage}{\linewidth}\noindent 
\ifpreprint
  \centerline{\epsfig{file=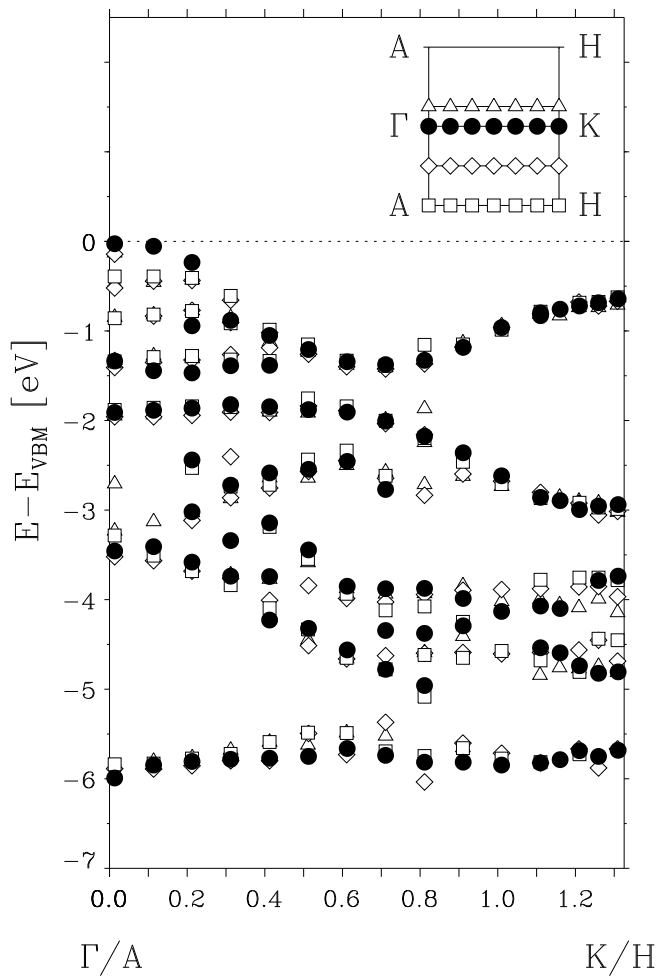, angle=0, width=0.8\linewidth}} 
\else
  \centerline{\epsfig{file=mos2_be.eps, angle=0, width=1.0\linewidth}} 
\fi
  \caption{The experimentally derived band structure of MoS$_2$ along 
    the $\Gamma$K and AH, as well as, two other parallel lines 
    as indicated in the inset. 
        \label{fig:all_gk}} 
\end{minipage} 
\end{figure} 

Fig.~\ref{fig:all_gk} presents the experimental band-structure features as
resulting for E($\vec{k}_\parallel$; $\vec{k}_\perp$ fixed) with the four
different $\vec{k}_\perp$ values noted above and indicated in the figure.
For the sake of clarity only pronounced emission features in the spectra 
are shown in the figure. The filled circles mark the $\Gamma$K 
direction with the $\vec{k}_\perp$ value of the $\Gamma$-point. In general, 
the symbols of the other three series of measured features for the other 
three $\vec{k}_\perp$ values lie very close to the circles. This is especially 
striking for the upper bands near K/H. Only the two uppermost measured bands
near $\Gamma$/A show a significant dependence on the $\vec{k}_\perp$ value. 
Also the bands between -2.0~eV and -6~eV do not reveal a very pronounced 
$E(\vec{k}_\perp)$ dependence. 

\begin{figure}\noindent 
\begin{minipage}{\linewidth}\noindent 
  \centerline{\epsfig{file=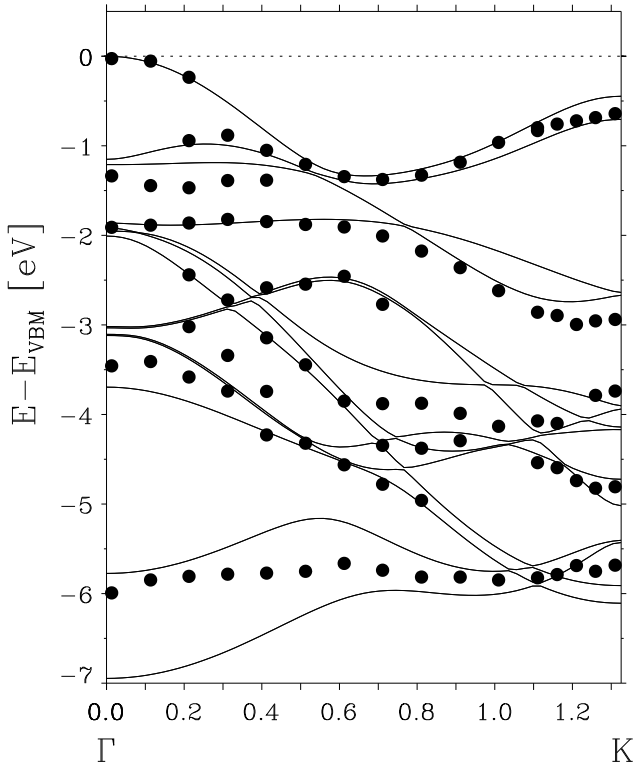, angle=0, width=1.0\linewidth}} 
  \caption{Experimentally derived band structure (symbols) of MoS$_2$ along 
    the $\Gamma$K line in comparison with the respective calculated bulk 
    band structure (solid lines). 
        \label{fig:gamma_gk}} 
\end{minipage} 
\end{figure} 

At this point it is very revealing to compare some of the above discussed experimental
results with results of our band-structure calculations. A respective comparison 
for $\vec{k}_\perp$ equivalent to zero, i.e., along the 
$\Gamma$K direction is shown in   
Fig.~\ref{fig:gamma_gk}. An amazingly good agreement between theory and 
experiment for most of the bands is obvious in Fig.~\ref{fig:gamma_gk} lending 
further support to the accurate experimental determination of the high symmetry line. 
Most of the Mo $4d$- and S $3p$-derived bands between 0 eV and -5 eV are
found to be in very good agreement. Only the S $3p$-derived band around -6 eV
shows significant deviations. One possible origin of this behaviour in experiment 
could be the large width in the underlying data, especially at $\Gamma$, so that 
the assignment of the measured structures to a well-defined binding energy is difficult. 
Another origin of the seeming disagreement could be related to the 
symmetry dependence of the observed spectral features.
We come back to this point in Section V B. The band in question originates from 
sulfur $3p$ valence states and its dispersion is determined by the 
interaction of these states across the van der Waals gap.

To obtain further evidence that the K and the M point at the zone edge can be 
accurately determined, CFS series along the KH and ML high-symmetry 
lines were recorded. The respective $\vec{k}_\perp$ values have been calculated
by using Eq.~\ref{eq:k_perp} even if this approximation is quite coarse. This is
sufficient since the determination of the critical points has been done experimentally.
In this case, MoSe$_2$ was chosen as the example since
it shows a distinct spin-orbit splitting along the KH line in addition to a 
clear three-dimensional band character. After analyzing
the spectra by the curve fitting approach discussed above, a weak 
$E(\vec{k}_\perp)$ dependence is observed in the data.

\begin{figure}\noindent 
\begin{minipage}{\linewidth}\noindent 
   \centerline{\epsfig{file=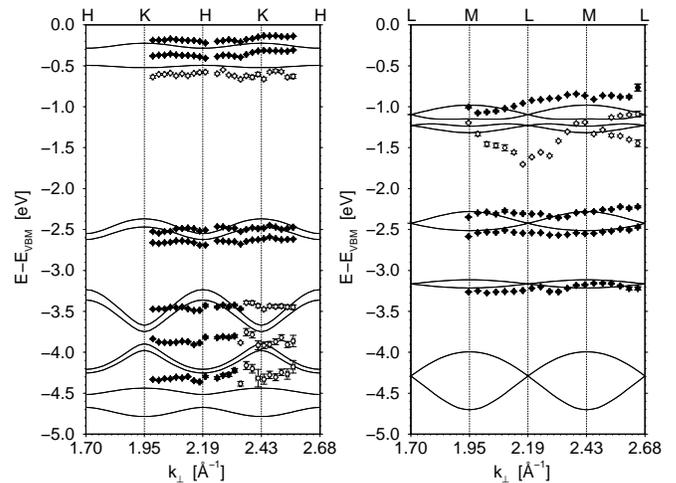, angle=-90, width=1.0\linewidth}} 
   \caption{Experimental valence-band structure (symbols) of MoSe$_2$ 
    along the symmetry lines KH and ML in comparison with the calculated 
    bands (solid lines).
        \label{fig:mose2_edge_fit}} 
\end{minipage} 
\end{figure} 

This is shown in Fig.~\ref{fig:mose2_edge_fit} where the experimental
data are compared to the corresponding calculated bands. The energy axis 
is aligned to the valence-band maximum which results from further 
investigations presented in the next sections. First we note that there is 
a reasonable general correspondence between the number and energy position
of measured and 
calculated bands. Even the dispersion of the theoretical and experimental 
results show good agreement in cases. The absolute binding energy for the 
topmost emission features, however, are slightly lower than the energies 
of the respective calculated bands. One can therefore conclude from the 
dispersion and energy position of the topmost bands and from the bands 
in the energy region around -2.5~eV that the K and the M points can be
determined. By having the three high symmetry points, namey $\Gamma$, K, and M,
one is able to scan all high-symmetry lines.

Despite a good correspondence of the states with smaller binding energies
up to -3.0~eV the states between -3.5 eV and -4.5 eV along the KH symmetry 
line show a fairly poor agreement with theory. In particular, the calculated 
bands show a significant $\vec{k}_\perp$ dispersion while experiment observes
nearly dispersionless states. We have observed the same behaviour already 
in Fig.~\ref{fig:gamma_gk}, where two bulk bands at about -6 eV and -7 eV 
were found while experiment observed only one very narrow band. We come back 
to this point further below.

For the ML direction a direct comparison of the bands in the latter energy 
region is not possible, since the respective bands could not be observed 
in experiment because of the limited monochromator energy range.
%

\subsection{Valence-band structure along $\Gamma$A}

Normal emission spectra for all three materials were recorded at various 
beamlines at HASYLAB or BESSY-I synchrotron sources, respectively, in the 
energy range of 13$\dots$30~eV. The energy axis is referred to the valence-band 
maximum which was determined in connection with the complete 
valence-band structure as described in the following sections.
To determine the band structure from the spectra series all spectra were
modeled after the removal of a small Shirley background \cite{Shirle72} by a 
sum of Gaussian and Lorentzian profiles. The $\vec{k}_\perp$ value within the 
Brillouin zone to which each peak has to be attributed was calculated from 
Eq.~\ref{eq:k_perp}. As pointed out above, in this limited energy range
of about 10$\dots$30~eV this equation can be used allowing for  
a sufficiently precise calculation of the $\vec{k}_\perp$ value for
each energetic peak position. For larger energies the deviation of the final 
states from free-electron-like behavior must be taken into account.
Here target current spectroscopy might be applied to first determine the 
coupling bands and then measure all photoemission spectra in the CFS mode 
with the final state belonging to the specific coupling band 
\cite{StStNi96,Stroco95,StStNi96a}.

Before addressing the observed band structures along $\Gamma$A in detail 
we note that only each second $\Gamma$ point is observed in experiment.
According to Ref. \onlinecite{PeLaJo85} this is due to the fact that in
non-symmorphic crystal structures the symmetry of the final state is not
always unique as it is for symmorphic structures. The $\Gamma{\rm A}$ direction 
$\Delta$ consists of the points
$\vec{k}=2\pi/c(0,0,\delta)$ with $0<\delta<\frac{1}{2}$.
$\Delta$ represents a group with a number of operations which are listed
for instance in the above reference. For example a plane wave of the form 
$\exp[i2\pi/c(0,0,\delta)(x,y,z)]$ is transformed by any operation from 
$\Delta$ to an irreducible representation and this plane wave 
belongs to e.g.\  $\Delta_1$. In contrast a plane wave 
$\exp[i2\pi/c(0,0,\delta+1)(x,y,z)]$ which is streched by a reciprocal 
lattice vector $2\pi/c(0,0,1)$ belongs to $\Delta_2$.
Obviously the symmetry depends on the length of the wave vector.
For a given final state symmetry this leads to the observed missing of each 
second $\Gamma$ point since each second initial state reveals an odd symmetry.
This fact has to be taken into account in the comparison to the band structure 
calculations.

The experimental band structures along $\overline{\Gamma {\rm A} \Gamma}$ are 
shown in Figs.~\ref{fig:mos2_ga},~\ref{fig:mose2_ga},~and~\ref{fig:mote2_ga}
together with the results of our band structure calculations. In one case
(Fig.~\ref{fig:mos2_ga}) we have included the density of states of the bands 
along the $\Gamma$A direction, as well.
We first address the experimental results. In general, for all three materials 
one can easily distinguish at least two dispersive structures and a number of
nearly non-dispersive bands. The uppermost valence band, as well as, the band 
with the largest binding energy reveal a significant $\vec{k}_\perp$-dependence. 
The width of the uppermost band between $\Gamma$ and A decreases from MoS$_2$ 
to MoTe$_2$, i.~e.,  the bands become more two-dimensional and less 
$\vec{k}_\perp$-dependent. This effect is not so strong for the band with the 
largest binding energy but also noticeable. From this result one may conclude 
that the latter might be predominantly of chalcogen $p_z$ orbital character, 
which is relatively strongly influenced by the following sandwich layer in the 
crystal. Indeed, our band structure calculations exhibit a contribution of the
chalcogen $p_z$ orbitals from 66\% to 60\% for the three chalcogenides to this 
state. In contrast, the influence of the next sandwich layer on the predominantly 
Mo $4d_{z^2}$-derived states (61\% to 68\% metal $d_{z^2}$ character) at the 
valence-band maximum decreases rapidly with the increasing size of the unit cell.

\begin{figure}\noindent 
\begin{minipage}{\linewidth}\noindent 
        \centerline{\epsfig{file=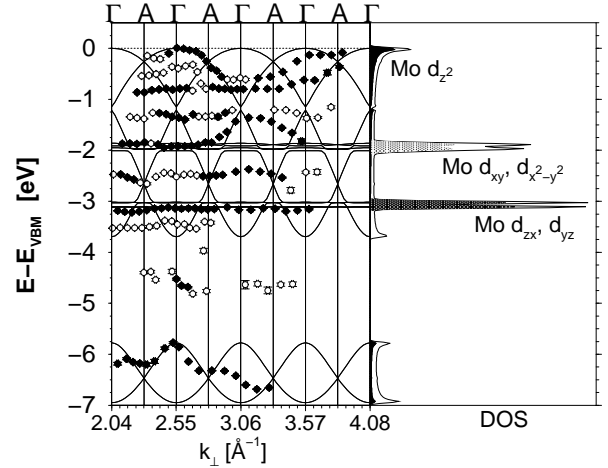, angle=0, width=0.9\linewidth}} 
        \caption[Experimental valence-band structure (symbols) of MoS$_2$ 
    along the $\Gamma$A high-symmetry line in comparison with the 
    calculated band structure (solid lines).]
    {Experimental valence-band structure (symbols) of MoS$_2$ 
    along the $\Gamma$A high-symmetry line in comparison with the 
    calculated band structure (solid lines). Open symbols denote weak  
    measured structures. The error bars are due to the limited experimental
    energy and angular resolution and due to uncertainties from the curve fitting.
    In addition, the density of states for the bands  
    along $\Gamma$A is plotted. The contributions of particular  
    molybdenum $d$-states is indicated by the shaded regions. 
        \label{fig:mos2_ga}} 
\end{minipage} 
\end{figure} 

\begin{figure}\noindent 
\begin{minipage}{\linewidth}\noindent 
        \centerline{\epsfig{file=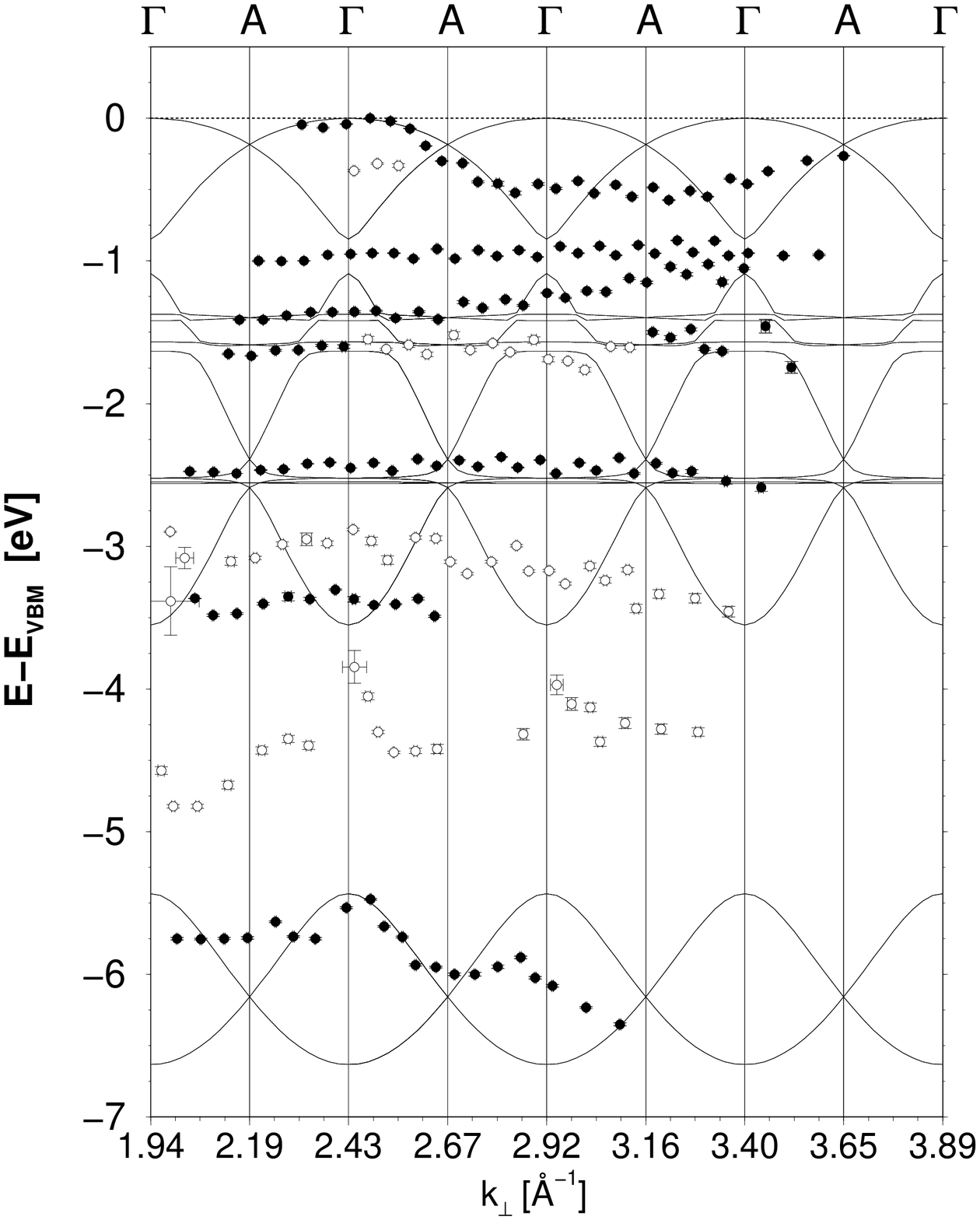, angle=0, width=0.8\linewidth}} 
        \caption[Experimental valence-band structure (symbols) of MoSe$_2$ 
    along the $\Gamma$A high-symmetry line in comparison with the 
    calculated band structure (solid lines).]
    {Experimental valence-band structure (symbols) of MoSe$_2$ 
    along the $\Gamma$A high-symmetry line in comparison with the 
    calculated band structure (solid lines). 
    Open symbols denote weak structures. The error bars are due to the limited experimental
    energy and angular resolution and due to uncertainties from the curve fitting.
        \label{fig:mose2_ga}} 
\end{minipage} 
\end{figure} 

\begin{figure}\noindent 
\begin{minipage}{\linewidth}\noindent 
        \centerline{\epsfig{file=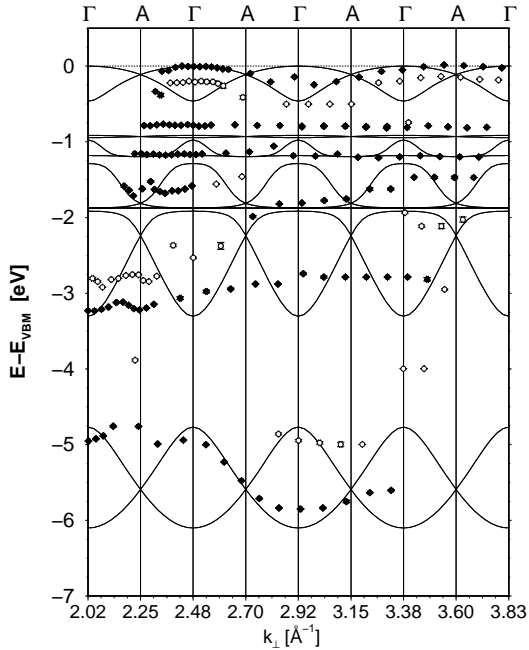, angle=0, width=0.8\linewidth}} 
        \caption[Experimental valence-band structure (symbols) of MoTe$_2$ 
    along the $\Gamma$A high-symmetry line in comparison with the 
    calculated band structure (solid lines).]
    {Experimental valence-band structure (symbols) of MoTe$_2$ 
    along the $\Gamma$A high-symmetry line in comparison with the 
    calculated band structure (solid lines). 
    Open symbols denote weak structures. The error bars are due to the limited experimental
    energy and angular resolution and due to uncertainties from the curve fitting. 
        \label{fig:mote2_ga}} 
\end{minipage} 
\end{figure} 

It should be noted that in all compounds the topmost band at the second 
$\Gamma$-point (for which it can be observed) lies at larger binding energies 
in comparison to the first one. This may be attributed to two effects. First, 
the analyzer has a fixed maximum acceptance angle $\vartheta_{max}$. The 
$\vec{k}$-resolution $\Delta k$ depends on this acceptance angle and on the 
kinetic energy (see Eq.~\ref{eq:k_par}). Assuming $\vartheta_{max}=0.8^\circ$ 
then for the first $\Gamma$-point $\Delta k\approx0.020\;{\rm [\AA^{-1}]}$ while 
for the second one it is  $\Delta k\approx0.036\;{\rm [\AA^{-1}]}$. Since the 
$\Gamma$-point is the point with the highest symmetry the band disperses in all 
$\vec{k}$-directions away from $\Gamma$ to larger binding energies. For a lower 
$\vec{k}$-resolution the band position at all $\vec{k}$ values within this range 
must be integrated. This yields an emission feature with a maximum at a larger 
binding energy. The second effect might be the reduced information depth at higher
photon energies. If the mean free path becomes lower than one unit
cell, the $\vec{k}_\perp$ dependence will smear out and the spectra
will reflect a more one-dimensional density of states \cite{FiHeSt97}. 
 
Below the two dispersive bands a number of nearly perfect dispersionless
bands are observed. The density of states resolved with respect to Mo $4d$ 
and S $3p$-character in Fig.~\ref{fig:mos2_ga} clearly reveals the physical
origin of some of these features. From the results of our band-structure calculations
it is most obvious, that these bands are two-dimensional and can be attributed
to Mo $4d$ states (see the calculated density of states in Fig.~\ref{fig:mos2_ga}
for that matter). In the calculations, they are located at about - 1.8~eV and 
- 3.1~eV and the bands are separated in both cases due to spin-orbit interaction. 
For MoS$_2$ this splitting is too small to be observed in experiment.
In contrast, for the other two compounds the photoemission data clearly exhibit 
the splitting of the dispersionless bands at about - 1.5\,eV and - 1.1 eV, 
respectively. In MoSe$_2$ this energetic separation is about 300~meV and for 
MoTe$_2$ it is about 500~meV. So, the energetic separation of these two spin-orbit 
splitted band groups, as well as,
their binding energy decrease towards the tellurium compound in very good 
agreement between theory and experiment. The density of states in the other 
energy regions (see Fig.~\ref{fig:mos2_ga}) is comparatively small, 
so that respective band-structure features are more complicated to measure.

Thus a comparison with the calculated band structures in general shows very 
good agreement, indeed, in spite of the fact that some experimentally 
observed emission features do not result from the calculations. In all 
three compounds such emissions are found above the lowest binding energy 
state at about -4.5\,eV. Typically this is only a very weak structure but 
it is also directly visible in the spectra of e.g.\  MoS$_2$ in the energy
range of 21$\dots$23~eV. Additional states may also arise from transitions 
of an initial state to a neighboring final state with different binding energy.
Since an energy distribution curve provides only combined information of 
the initial, as well as, the final state it cannot be distinguished between 
such effects. Of course, one would expect that only one coupling band is suitable 
for a given initial state but as it has been stated by Strocov {\it et al.} 
\cite{StStNi98}, in some 
cases an additional final state branch may exist. The emission intensity is then 
expected to be lower as from the main branch and this can be supported by the 
observed features in the present spectra.

Additional valence bands between the non-dispersive bands and the topmost valence 
band are experimentally observed but not supported by theory. A number of the 
respective spectral features are quite intense and might not be explained 
by different coupling branches. In MoS$_2$ at least two strong emission 
features appear at about -1.3~eV and -0.8~eV which are quite
non-dispersive for lower photon energies. For higher photon energies, i.~e.,
beyond the second $\Gamma$ point the band near -0.8~eV begins to disperse 
to smaller binding energies while the band which is found energetically 
below the former vanishes.

Finally, we note that all three valence band structures in 
Figs.~\ref{fig:mos2_ga},~\ref{fig:mose2_ga},~and~\ref{fig:mote2_ga}
show valence bands within the energy region from - 5 eV to - 7 eV
with significant $\vec{k}_\perp$ dispersion. This dispersion shows the 
symmetry properties, as discussed above, in that only one of the two 
bands between consecutive $\Gamma$-points is seen in experiment. The 
agreement between the respective theoretical and experimental bands 
is particularly close for MoTe$_2$ as can be seen in Fig.~\ref{fig:mote2_ga}.
By the same token, it becomes most obvious, why experiment observes in 
this energy region for one particular $\Gamma$ point either the upper or 
the lower band but not both. Since the experimental data plotted in 
Fig.~\ref{fig:gamma_gk} were recorded for one particular $\vec{k}_\perp$ 
at the $\Gamma$-point only one band could be observed.


\subsection{Valence-band structure along the in-plane directions}
\label{help}

After the discussion of the valence-band structure in the
$\Gamma {\rm A}$ direction we now focus on the directions parallel to the 
layers. We discuss the band structures along the high-symmetry lines $\Gamma 
{\rm K}$, $\Gamma {\rm M}$, and MK of the Brillouin zone. 
The $\Gamma {\rm K}$ direction is the longest high-symmetry line. In the case 
of the molybdenum dichalcogenides this direction is most important since the 
energetic position of the topmost emission features at the $\Gamma$ and the 
K point determine a number of electrical and optical properties. 

\ifpreprint
\else 
        \end{multicols} 
\fi 

\begin{figure}\noindent 
\begin{minipage}{\linewidth}\noindent 
        \centerline{\epsfig{file=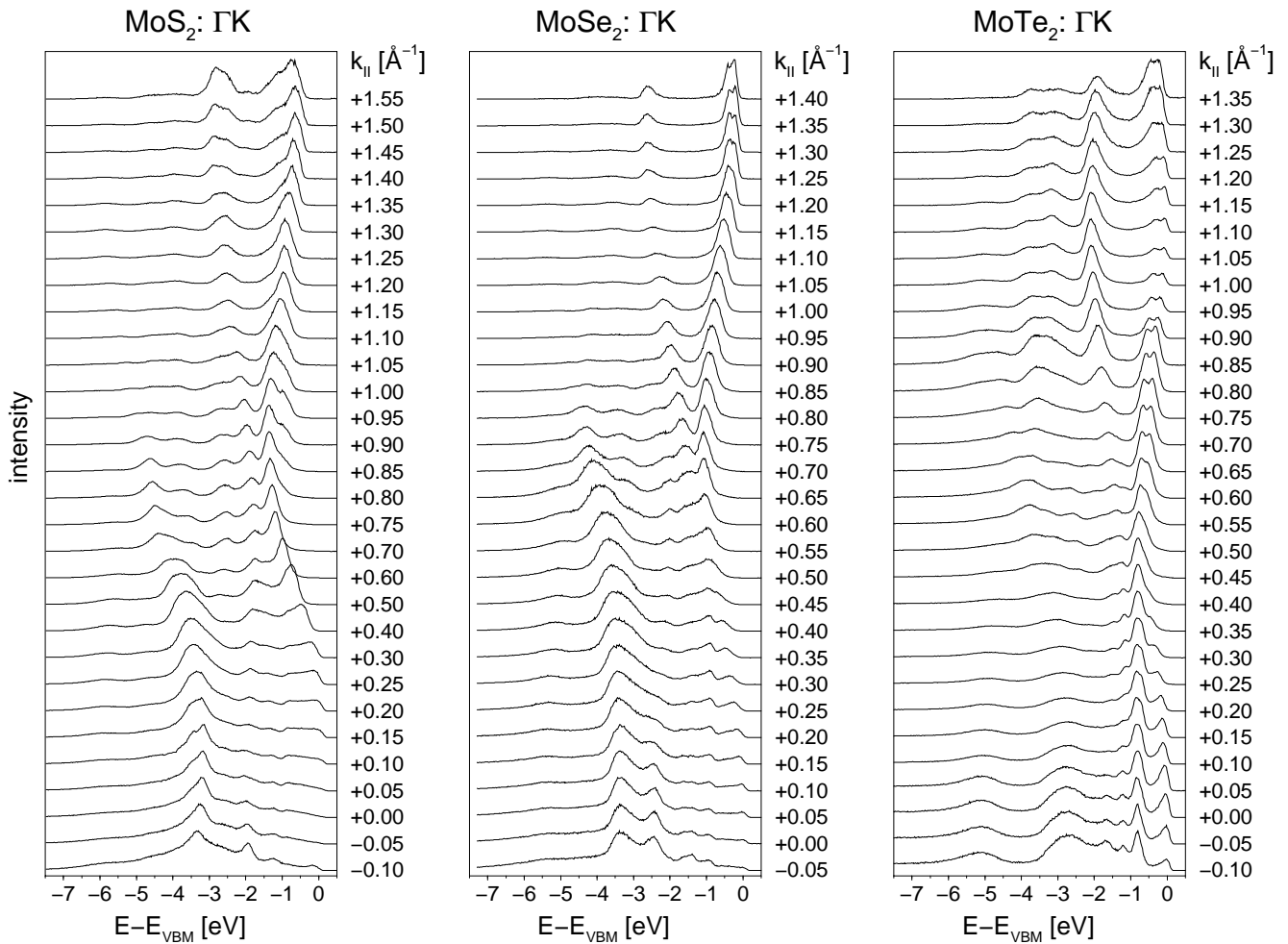, angle=270, width=1.0\linewidth}}  
        \caption[ARPES spectra along the high-symmetry line $\Gamma {\rm K}$ 
        of the three molybdenum dichalcogenide compounds.]{ARPES spectra 
        recorded in the CFS mode along the high-symmetry  
    line $\Gamma {\rm K}$ of the three molybdenum dichalcogenide compounds.  
    The size of $k_\parallel$ at the K-point is given in Tab. \ref{tab:strcdata}. 
        \label{fig:GK_series}}  
\end{minipage} 
\end{figure} 

\ifpreprint 
\else 
        \begin{multicols}{2}
\fi 

Fig.~\ref{fig:GK_series} shows CFS spectra for the three molybdenum 
dichalcogenides recorded along the $\Gamma {\rm K}$ line. 
The respective final state was chosen such that $\vec{k}_\perp$ remains 
in the $\Gamma$KM plane within the whole series of spectra. Here the 
experimental determination of the $\Gamma$, K, and M point has been taken
into account.

A number of features is easily recognized in the spectra of
Fig.~\ref{fig:GK_series}. The three series reveal a close
similarity of the $E(\vec{k})$ behavior of most emission features.
Starting at $\Gamma$ (i.~e.,  $k_\parallel=0.0\; {\rm \AA^{-1}}$)
to the K point the states near -3.2~eV and the lowest binding energy
disperse towards larger binding energies.
This is different from the two topmost emission features,
which separate from each other after nearly halfway towards K
and move with nearly constant width to K.

From the spectra in Fig.~\ref{fig:GK_series} it can be deduced 
that for MoS$_2$ the topmost emission feature near $\Gamma$ reveals the 
weakest intensity compared to the other compounds. Since this state is 
predominantly derived from the Mo $4d_{z^2}$ orbitals (see previous
section) high spectral intensity for this state is expected. It should be 
decreased for a lower lattice parameter $a$ due to a shielding 
by the chalcogen $p$ orbitals. Indeed this is observed. Furthermore, 
the topmost emission feature reveals a decreasing
dispersion from MoS$_2$ to MoTe$_2$. 

The spectra for the $\Gamma {\rm K}$ and $\Gamma {\rm M}$ directions are 
largely similar. The spectra along the MK direction show relatively small 
dispersions, as compared to those along $\Gamma$K and $\Gamma$M.
Only the features with the smallest binding energy show appreciable dispersion.


\subsection{The spin-orbit splitting at K}
Before discussing the complete valence-band structure of the three molybdenum 
dichalcogenides, we first briefly address the effects of spin-orbit 
coupling on the spectra and the band structures. A distinct splitting of the 
emission features with lowest binding energy at the K point can be observed 
in Fig.~\ref{fig:GK_series}. It is most obvious for MoTe$_2$ and is only 
visible as a weak shoulder in the MoS$_2$ compound. Fig.~\ref{fig:K_spect} 
shows the relevant part of the spectra for each compound where the double 
emission feature is assigned with X and Y, respectively, in line with previous 
work on MoTe$_2$ \cite{BoMuAu99}. A broad additional emission feature denoted 
as `add.' is taken into account which is present in all three spectra just 
below peak X. 

\begin{figure}\noindent 
\begin{minipage}{\linewidth}\noindent 
\ifpreprint
    \centerline{\epsfig{file=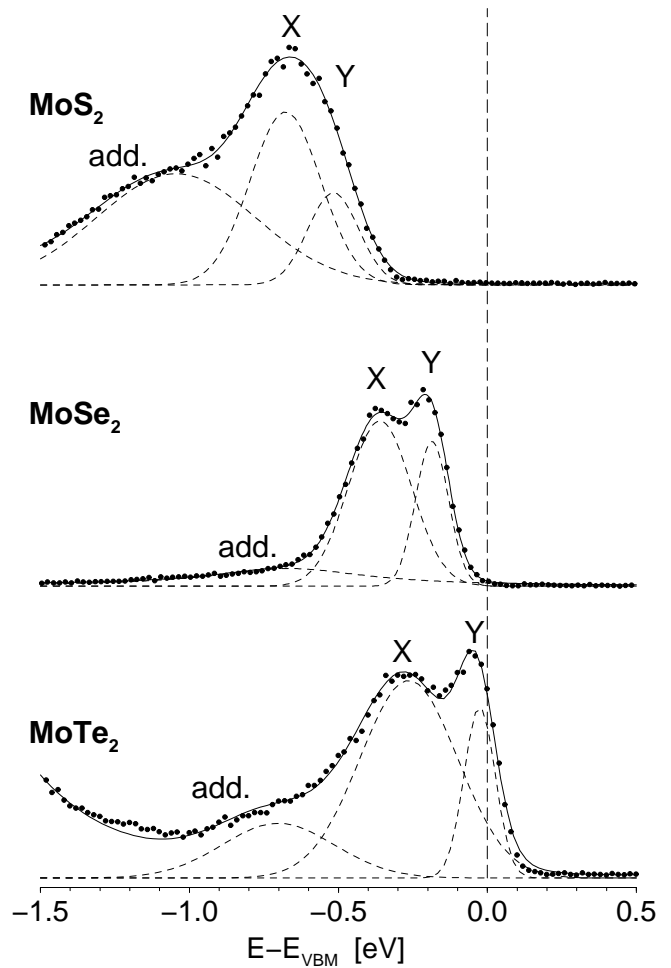, angle=0, width=0.80\linewidth}} 
\else
    \centerline{\epsfig{file=k_spekt.eps, angle=0, width=1.00\linewidth}} 
\fi
    \caption[Double emission feature for the three dichalcogenides at the K point.]
   {Double emission feature for the three dichalcogenides at the K point.  
   Filled circles denote the experimental data while the solid lines represents the  
   fit curve as obtained from the sum of two Gaussians (dashed lines) 
    (X, Y) and an additional weak emission feature labelled "'add."'. 
        \label{fig:K_spect}} 
\end{minipage} 
\end{figure} 

The size of the splitting of these bands at the K-point in the experimental 
spectra and in the theoretical results is listed in Tab.~\ref{tab:K_split}. From
MoS$_2$ to MoTe$_2$, the splitting increases with increasing Mo $d$ character 
of the bands. Furthermore, the energetic position of the bands move towards the
valence-band maximum at the $\Gamma$-point (cf. Fig.~\ref{fig:K_spect}). 

The general trend in the observed energetic differences whithin the row of the 
three dichalcogenides is observed in the theoretical results, as well. The 
experimentally observed splittings of the emission features X and Y at the 
K-point, however, are about 100 meV smaller than the respective calculated 
splittings. 

Another point concerning the splitting at K is connected with
the interpretation of two structures in optical measurements.
Optical absorption spectra of all three MoX$_2$ compounds display 
excitonic features characteristic of the group VIA transition metal
dichalcogenides \cite{BeKnLi72,BeaHug79}.
As claimed by Coehoorn {\em et al.}\/ \cite{CoHaGr87}, who observed a
splitting of 0.21~eV for MoSe$_2$ at the K
point, these states give rise to the so-called A and B excitons. The 
splitting found in Ref. \onlinecite{CoHaGr87} for MoSe$_2$ matches well 
with the observed energetic difference of the A and B excitons (see, e.g.,  
Ref. \onlinecite{BeLiHu76}) and therefore the authors concluded that probably 
the exciton peaks arise from the two possible transitions at the K point.

\begin{minipage}{\linewidth}\noindent 
\begin{table}[hbt] 
 \caption[Energetic separation of the emission features X and Y at the  
        K point in comparison with the calculated splitting.]{Energetic separation of the emission features X and Y at the  
        K point in comparison with the calculated splitting. 
        The right column gives the energetic distance between the A and B 
        excitons following from optical absorption 
        measurements~\cite{BeKnLi72}.} 
\begin{tabular}{l|cc|c} 
  & \multicolumn{2}{c|}{Splitting at K} & A, B excitons     \\   
  & Exp.\  ARPES data &  Calculation & Exp.\  opt.\  data \\  \hline 
MoS$_2$   & $161\pm10$~meV    &  258~meV &    180~meV    \\ 
MoSe$_2$  & $175\pm10$~meV    &  294~meV &    250~meV    \\ 
MoTe$_2$  & $238\pm10$~meV    &  331~meV &    380~meV    \\ 
\end{tabular} 
\label{tab:K_split} 
\end{table} 
\end{minipage} 

Additionly, Finteis {\it et al.}\/ found the valence-band maximum of
WSe$_2$ to be located at the K point in their experiments \cite{FiHeSt97,FiHeSt99}.
The latter authors also concluded that the double emission feature at the 
K-point is the signature of the two initial states of the excitons A and B.
To attribute the double emission feature at the K point to the initial
states of the excitons A and B is at variance, however, with several investigations
using optical absorption measurements \cite{BeKnLi72,BrMuYo72,TaFuKu78}
leading to the result that appropriate transitions for the A and B
excitons should occur at the $\Gamma$ point. However, according to the 
present experimental data the peak distances as listed
in Tab.~\ref{tab:K_split} for the double feature X, Y for all three
materials are too small compared to the energetic 
distance of the excitonic lines measured by e.g.\  Beal {\em et al.}\/ 
\cite{BeKnLi72}. Even though a shift between the initial states and the 
observed energy distance of the excitons is possible \cite{OnoToy67}
the magnitude of the observed shift cannot be explained within the
theory \cite{OnoToy67}. It must be concluded accordingly that the double 
emission feature X, Y at the K point does not correspond to initial states 
of the excitons A and B \cite{BoMuAu99}. In line with the optical absorption 
measurements cited above and from the location of the VBM at the $\Gamma$ 
point it is suggested that the topmost occupied state at $\Gamma$ is the 
initial state for the A and B excitons. 


\subsection{Complete valence-band structures}

In Figs.~\ref{fig:mos2_kp},~\ref{fig:mose2_kp}, and~\ref{fig:mote2_kp}
the experimentally observed spectral features (see section \ref{help})
of each molybdenum dichalcogenide compound are plotted as a band structure 
along the high-symmetry lines $\rm M\Gamma KM$ in direct comparison 
with the results of our band-structure calculations. A projected band structure has
not been used since the respective critical points have been determined 
experimentally.

The lower unoccupied bands are also shown. Except for the gap problem
within the LDA the dispersion of these bands should be correct.
The comparison with the results of inverse photoemission investigations 
on MoS$_2$  \cite{LaBeTh95,SaBrCh88} confirms this finding.
Also studies on the similar material WSe$_2$ show good qualitative
agreement \cite{VoKrMaPo99}.
Thus it can be concluded that the minimum of the conduction bands is 
located nearly halfway between $\Gamma$ and K, indeed. Furthermore, 
the valence-band maximum is found at $\Gamma$ in all three compounds.
This is confirmed for all three materials by the experimental data. From 
the present band structure calculations it follows that 
the energetically lowest indirect transition is from $\Gamma$ 
to $\approx \frac{1}{2}\Gamma$K and the lowest direct transition 
occurs at K. 

A close look at the valence-band structure of all three materials
reveals a very good general agreement between measured and calculated
bands. The total valence-band width is
found to decrease from MoS$_2$ over MoSe$_2$ to MoTe$_2$ both in  
experiment and theory.

The dispersion of the uppermost bands away from $\Gamma$ towards K and M
reveal also quantitative matching. However, the energetic positions of
the calculated bands in the vicinity of the high-symmetry point K
differ with a maximum difference of 200~meV in the case of MoSe$_2$.
Also the calculated splitting of the bands at K (already discussed above)
is about 100 meV larger than observed in experiment. The general trend in 
decreasing binding energy of the top of the valence bands at M, as observed 
in experiment, results from theory, as well. Only the absolute energetic 
location of the topmost state at M is about 200~meV too low as compared to the 
photoemission results. It must be emphasized that such differences between 
experiment and theory in the range of a few 100~meV are close to the theoretical
uncertainty that can be achieved nowadays. 

In the vicinity of about -1~eV a dispersionless band is observed in the 
$\Gamma$A direction (see Figs.~\ref{fig:mos2_ga},~\ref{fig:mose2_ga},
and~\ref{fig:mote2_ga}) which is also dispersionless near the center of the 
Brillouin zone for off-normal emission (see Figs.~\ref{fig:mos2_kp},~\ref{fig:mose2_kp},
and~\ref{fig:mote2_kp})~\cite{kom}. In the $\Gamma$KM-plane, this band is found 
experimentally in MoS$_2$ to be $\approx$100~meV below a rather dispersionless 
band in the calculated band structure and about 100~meV above the respective 
band for MoSe$_2$ and MoTe$_2$. On the other hand, along $\Gamma$A only in the 
case of MoTe$_2$ this band results from the theory.

In band structure calculations for the (0001)-surface of MoX$_2$ (X=S, Se, Te), 
which we have carried out using a slab of 9 sandwich layers (method as 
described in~\cite{VoKrMaPo99}), we have found states located at the first few 
sandwich layers of the surface in this energy range. The respective bands exhibit 
only a small dispersion around the center of the surface Brillouin zone in 
agreement with the experimentally observed bands. Therefore, one might attribute 
these bands in the photoemission data to surface states.

\begin{figure}\noindent 
\begin{minipage}{\linewidth}\noindent 
        \centerline{\epsfig{file=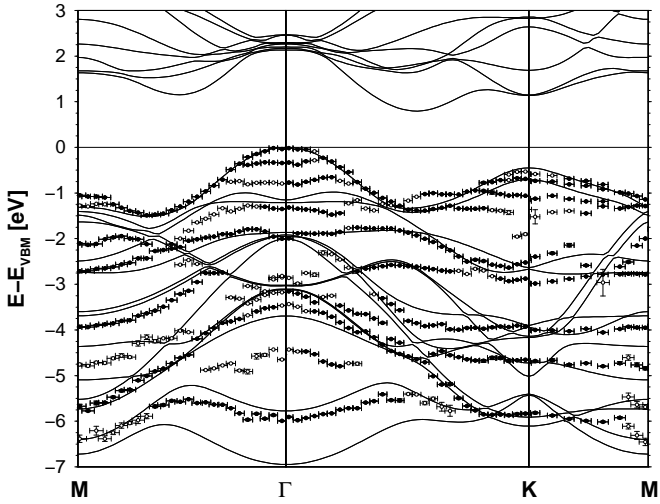, angle=270, width=\linewidth}} 
        \caption[Experimental valence-band structure (symbols) of MoS$_2$  
    along the in-plane directions.]{Experimental valence-band structure (symbols) of MoS$_2$  
    along the in-plane directions M$\Gamma$, $\Gamma$K, and KM in 
    comparison with the calculated band structure (solid lines). 
    Open symbols denote weak structures. The error bars are due to the limited experimental
    energy and angular resolution and due to uncertainties from the curve fitting.
        \label{fig:mos2_kp}} 
\end{minipage} 
\end{figure} 
 
\begin{figure}\noindent 
\begin{minipage}{\linewidth}\noindent 
        \centerline{\epsfig{file=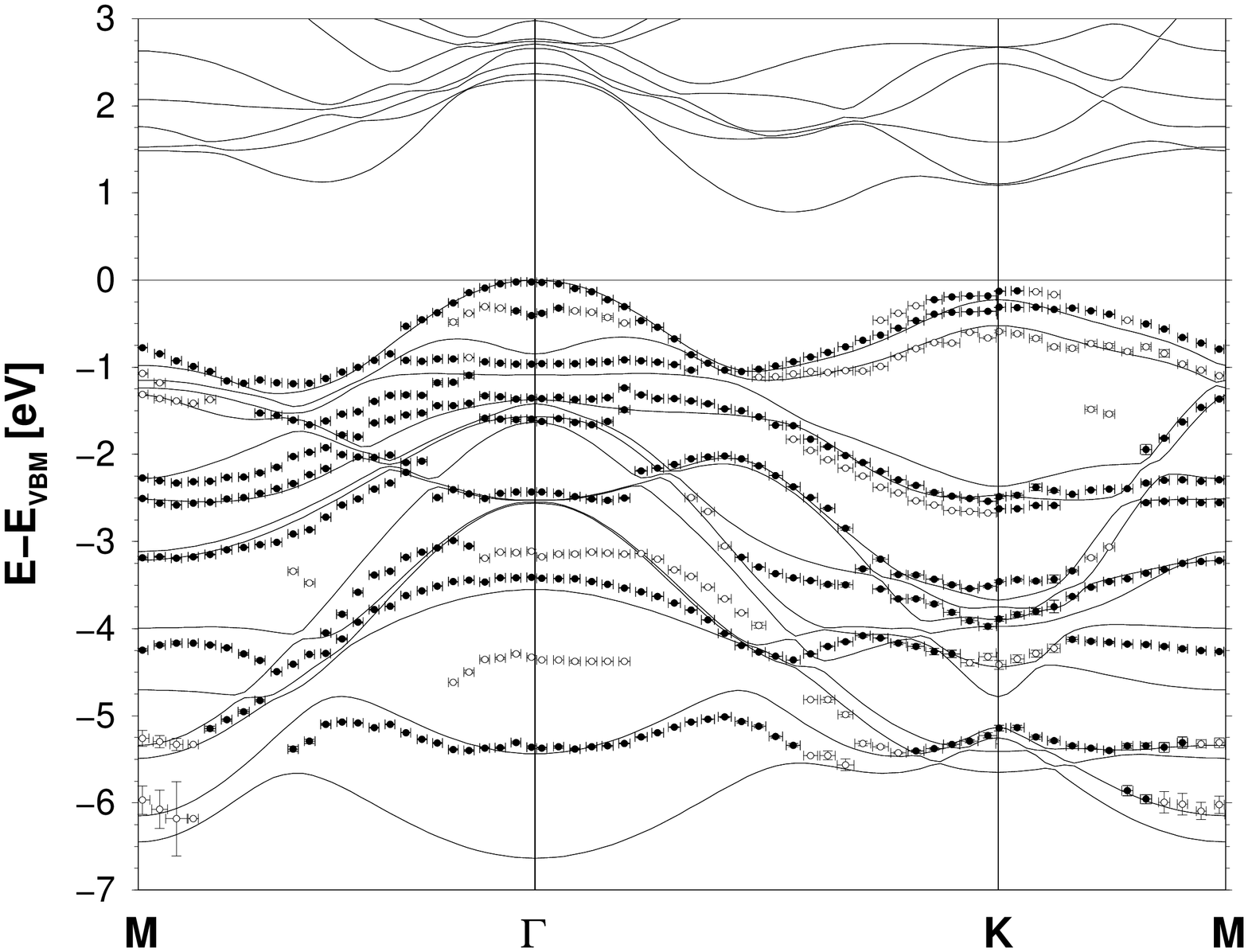, angle=270, width=\linewidth}} 
        \caption[Experimental valence-band structure (symbols) of MoSe$_2$  
    along the in-plane directions.]{Experimental valence-band structure (symbols) of MoSe$_2$  
    along the in-plane directions M$\Gamma$, $\Gamma$K, and KM in 
    comparison with the calculated band structure (solid lines). 
    Open symbols denote weak structures. The error bars are due to the limited experimental
    energy and angular resolution and due to uncertainties from the curve fitting. 
        \label{fig:mose2_kp}} 
\end{minipage} 
\end{figure} 
 
\begin{figure}\noindent 
\begin{minipage}{\linewidth}\noindent 
        \centerline{\epsfig{file=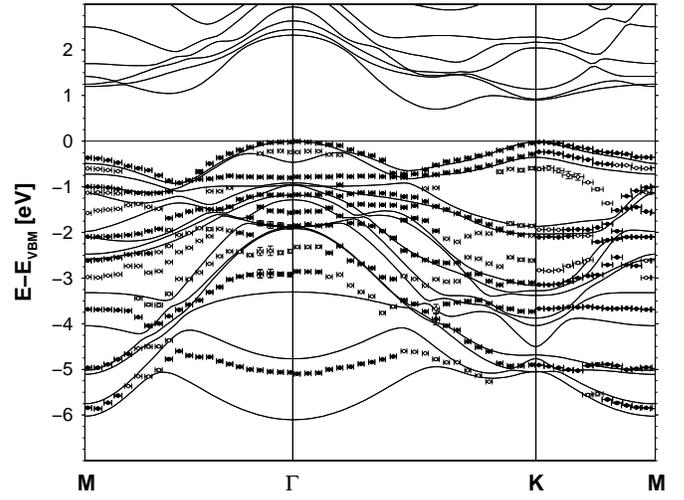, angle=270, width=\linewidth}} 
        \caption[Experimental valence-band structure (symbols) of MoTe$_2$  
    along the in-plane directions.]{Experimental valence-band structure (symbols) of MoTe$_2$  
    along the in-plane directions M$\Gamma$, $\Gamma$K, and KM in 
    comparison with the calculated band structure (solid lines). 
    Open symbols denote weak structures. The error bars are due to the limited experimental
    energy and angular resolution and due to uncertainties from the curve fitting.
        \label{fig:mote2_kp}} 
\end{minipage} 
\end{figure}

\section{Conclusions}

In our investigations, we have determined the complete valence-band
structure of MoS$_2$, MoSe$_2$, and $\alpha$-MoTe$_2$ on equal footing
by experiment and theory. It was found that the valence bands
resulting from angle-resolved photoelectron spectroscopy in the 
constant-final-state mode and from {\em ab-initio}\/ calculations
are in very good overall agreement. Our results yield the first comprehensive 
picture of the electronic bulk valence-band structure of all three 
molybdenum dichalcogenides.

\section{Acknowledgement}

We would like to thank D.~Kaiser for natural MoS$_2$ samples, A.~Klein from 
Technische Universit{\"a}t Darmstadt for the MoSe$_2$ samples, and
D.~Kaiser and J.~Augustin for the MoTe$_2$ samples.
This work received funding from the Bundesministerium f\"ur Bildung, 
Wissenschaft, Forschung und Technologie (BMBF) under
project \#05 622 KHA and \#05 SE8 PMA 6.



\ifpreprint
\else 
        \end{multicols} 
\fi



\begin{references}
\bibitem{kom}
Just below -1~eV a rather dispersionless emission feature 
is observed for all three compounds from $\Gamma$ to about halfway to K.
For MoSe$_2$ an equivalent, rather dispersionless state was reported 
\cite{CoHaDi87} and discussed in the context of a vibronic polaron
similar to the electronic polaron claimed for TiTe$_2$ \cite{BoBrBu84}.
Recent studies \cite{ClAnGw96} on TiTe$_2$ disprove the electronic
polaron concept so that also in the case of similar layered transition 
metal dichalcogenides other explanations may be more plausible.
The respective peak in the $\Gamma {\rm K}$ series might be
non-dispersive simply due to a weak overlap with other mainly chalcogen
derived orbitals in the direction close to the z-axis.
With increasing off-normal emission 
angles $\vartheta$, i.e.~reaching the Brillouin zone edge, the small 
overlap with the chalcogen $p_z$ orbitals gives rise to the observed
dispersion.
\end{references}
\end{document}